\documentstyle[12pt]{article}

\newcommand{\be}{\begin{equation}}
\newcommand{\ee}{\end{equation}}

\begin{document}

\begin{center}

{\Large {\bf Critical behavior of nonequilibrium $q$-state systems}}

\vspace{1.cm}

{\bf Andrea Crisanti$^1$ and Peter Grassberger$^2$}

$^1$ Dipartimento di Fisica, Universit\`a di Roma I, Ple. Aldo Moro,
    Roma, Italy

and

$^2$ Physics Department, University of Wuppertal, D-42097 Wuppertal, FRG

\vspace{.8cm}

\today

\vspace{1.1cm}

{\bf Abstract} \end{center}

{\small \advance \baselineskip by -2pt

We present two classes of nonequilibrium models with critical
behavior. Each model is characterized by an integer $q>1$, and is
defined on configurations of $q$-valued spins on regular lattices.
The definitions of the models are very similar to the updating
rules in Wolff's algorithm for the Potts model, but both classes
break detailed balance, except for $q=2$ and $q=\infty$. In the
first case both models reduce to the Ising model, while one of
them reduces to percolation (more precisely, to the general
epidemic process) for $q=\infty$. Locations of the critical point
and critical exponents are estimated in 2 dimensions.
%For the first model class, they interpolate monotonically between
%the Ising and percolation limits.
}


\eject

It is by now well understood  that nonequilibrium models can show
critical behavior very similar to second order phase transitions
in equilibrium systems. Indeed, dropping the requirement
of detailed balance gives additional freedom, and one can
observe a much richer spectrum of possibilities. These include
in particular critical behavior for generic values of the control
parameters, i.e. systems which do not require the control parameter
to be set to any non-natural value to be critical. Examples for
this are diffusion limited aggregation \cite{witten,meakin} and
various models of self-organized criticality \cite{bak,langer,drossel}.

A problem which is more common in nonequilibrium models than
in equilibrium cases is that it is often not clear how large are
the universality classes. Thus, e.g., it is not clear yet whether
different versions of the sand pile model \cite{bak,manna} are
in the same universality class or not. An other much discussed
example is provided by models with an absorbing state. The
prototype of such models is directed percolation, interpreted as
an epidemic process without immunization. By now it
seems clear that a model for heterogeneous catalysis proposed some
years ago by Ziff {\it et al} \cite{zbg} does belong to the same
universality class, in spite of original numerical indications of
the contrary \cite{jensen}. But there exist still a number
of similar models (some with degenerate absorbing states) for
which conflicting evidence has been reported as concerns universality.

Another problem which is more frequent in nonequilibrium cases is
that there exists no good field theoretic treatment. The best known
example for this is diffusion limited aggregation which is still not
amenable to such a treatment. This does not mean that one cannot
write down ``hamiltonians" \cite{doi,scheun} or path integrals
\cite{peliti}, but rather that no good perturbation expansions
are available which then could be resummed by renormalization group
methods. Other models for which this applies are the sand pile model
\cite{obukhov} and ``true" self avoiding walks \cite{amit,peli-pietro}
for which the field theories are intrinsically non-renormalizable
\cite{finnish}.

In the present note we want to present two new classes of nonequilibrium
critical phenomena. They are superficially similar to the Potts model
\cite{wu}, but they are not defined in terms of hamiltonians.
Instead, they are defined algorithmically, by stating the rules for
updating the state of a system. This is similar e.g. to the sand pile
model. But in contrast to that, our models show ``conventional"
critical behavior in the sense that they involve control parameters,
the critical behavior being observed only for special values. On the
other hand, we have not been able yet to apply field theoretic methods
to them.

Both class of models involve ``spins" with $q$ possible orientations
(``colors"), and are symmetric under cyclic permutations of the colors.
In addition, the first model is invariant under the exchange of any
two colors, just as the Potts model is.
The rules for updating a configuration are very similar to Wolff's
single cluster variant \cite{wolff} of the Swendsen-Wang dynamics
\cite{swendsen} of the Potts model. Let us thus recall Wolff's
algorithm as applied to the $q$-state Pots model.

Assume we have a configuration of spins on a hypercubic $d$-dimensional
lattice, with spin values ${s_i,\;i\in Z^d}$. The evolution is defined
by sequential updatings of randomly selected clusters. To make one
update, we proceed as follows: \\
(1) We pick a random color $s\in [0,1,\ldots q-1]$. \\
(2) Pick a random site $i\in Z^d$.
If $s_i=s$, we do not do anything and proceed to the next update. \\
(3) Otherwise, we build a {\it bond} percolation cluster on the subset
of sites $j$ which have $s_j=s_i$ and which are connected to $i$. \\
(4) After this is done, we flip this cluster, i.e. we change all its
$s_j$ to $s$, and proceed to the next update.

The control parameter in this model is the probability $p$ for
connecting sites in the percolation process in step (3). It is
related to the interaction strength $K$ in the Potts model, $e^{-\beta
H} = \exp(K\sum_{\langle i,j\rangle} (\delta_{s_i,s_j}-1))$, by
$p=1-e^{-K}$ \cite{swendsen}.

Our first class of models (``class A") is obtained from this
algorithm by keeping steps (1), (2) and (4), but replacing step (3)
by

(3A) Build a bond percolation cluster on the sites $j$ which are
connected with site $i$ and which have $s_j\neq s$.

Again this is done with probability $p$ for each bond. But for this
model we were not able to prove detailed balance for general values of
$q$, and we thus cannot
relate $p$ with a coupling strength in any hamiltonian.

In the second class of models (``class B") we skip step (1), keep
step (3), and replace the other steps by

(2B) Pick a random site $i\in Z^d$. \\
(4B) Flip all spins of the percolation cluster into color $s_i+1\;
{\rm mod}\, q$, i.e. rotate the entire cluster in color space by an
angle $2\pi/q$.

In this model the braking of detailed balance is evident.

The simplest case is $q=2$. In this case, the Potts model goes over
into the Ising model if we perform the trivial relabeling $s=0 \to
s=-1$. It is easy to see that in this case also models A and B are
equivalent to the Ising model.

Another simple limit is $q=\infty$. In this limit, $s_j$ will be
different from $s$ with probability 1, and thus the cluster built
in step (3A) is just an ordinary bond percolation cluster on the
entire lattice. Thus model A is equivalent to the spreading of
bond percolation according to the general epidemic process \cite{epid}
or the Leath algorithm \cite{leath}.

The $q\to\infty$ limit of model B is rather different. It is easy to
see that there is no ordering for any finite $p$ in this limit. For
finite $q$ and $p$, there is a balance between ordering due to the fact
that two sites $i,k$ which had originally $s_j=s_k-1$ might get the
same colors by flipping $j$, and disordering because a coherent cluster
is broken up. For $q\to\infty$, the ordering disappears, and thus
the transition point has to shift to $p_c\to 1$. For $q=\infty$ all
clusters are trivial (just one site) for all $p$, suggesting that the
transition turns into first order (all scaling law amplitudes $\to 0$)
for $q\to\infty$. Notice that this conclusion
rests on the fact that the fractal dimension of percolation
clusters at $p\leq p_c$ is zero, whence the ordering by each cluster
flip effects only a vanishing fraction of sites, and a random initial
configuration remains essentially random. This is not so on finite
lattices (in particular at $d=2$), since there the codimension of
percolation clusters is very small, and hence typical cluster flips
involve large parts of lattice. Thus we expect very large finite
size effects in model B even for finite values of $q$.

Let us next discuss mean field theory. In both models, this implies
that a spin not yet checked during the build-up of a cluster has the
same probability $1/q$ to have any of the $q$ possible colors. In a
strict mean field treatment one would assume the same also for sites
which had already been visited during the present cluster evolution.
We shall in the following discuss a more realistic ``hybrid" mean
field ansatz where we do not make the latter approximation. In this
case both models reduce to mixed site-bond percolation. In model A
we have bonds established with probability $p$, and sites with
probability $1-1/q$. In model B, the bond probability is the same,
but the site probability is $1/q$. This implies correctly that model
A turns into bond percolation for $q\to\infty$. But it predicts a
qualitatively wrong behavior for model B, as it would imply that model
B does not become critical for $q>1/p_{c,site}$, where $p_{c,site}$
is the threshold for site percolation on the same lattice. This is
obviously a weakness of our mean field ansatz, as it would mean that
even the Ising model ($q=2$) is non-critical in 2 dimensions. The
same problem would occur for the strict mean field treatment mentioned
above, and for the original Swensen-Wang-Wolff model with $q>2$.

In following we shall report results from simulations on 2-d square
lattices of size $N\times N$ with periodic boundary conditions. We
used both depth-first \cite{tarjan,wang} and width-first (``Leath"
\cite{leath}) algorithms for building the percolation clusters.
While the first are somewhat simpler when implemented by recursive
function calls, the temporal behavior is more natural in the latter
as it corresponds to epidemic like spreading \cite{epid}. The critical
point and the distribution of cluster sizes (and of cluster radii) is
identical with both algorithms. In all cases a sufficient number of
cluster flips during the initial transient stage was discarded, i.e.
all data reported below refer to the stationary state.

In order to obtain $p_c$ for model class A, we used two different
procedures. In the first (finite-size scaling) we used relatively small
lattices (up to $N=1024$), and determined $p_{c,N}$ from the requirement
that the cluster size distribution $P_N(s)$ shows the longest scaling
region for this value of $p$. In fig.1 this is illustrated for $q=3$
and $N=1024$. Values of $p_{c,N}$ resulting from this and analogous
plots are plotted against $1/N$ in fig.2. We see a straight line for
$N\geq 200$ which extrapolates to
\be
   p_c = 0.5330 \pm 0.0004
\ee
In the second procedure, we used much larger lattices (up to $8192\times
8192$) and sufficiently small values of $p$ so that all cluster
diameters were $<<N$ and finite-size effects were negligible. In fig.3,
resulting average cluster sizes $\langle s\rangle$ are plotted against
$p_0-p$ for three different values of $p_0$. Since we expect
\be
   \langle s\rangle \sim (p_c-p)^{-\gamma}.
\ee
we expect a straight line at $p_0=p_c$. This is indeed observed, with
the same $p_c$ as above and with $\gamma = 1.65\pm 0.05$. The error
is mainly systematic due to a slight but significant curvature in fig.3.

Similar deviations form scaling behavior are seen in the average cluster
evolution times and in the size distribution $P(s)$. The former
satisfies
\be
   \langle T\rangle \sim (p_c-p)^{-\delta}\;,\quad \delta \approx 0.85
\ee
(see fig.4), but there are rather strong deviations from a pure
scaling law.

{}From fig.1 we see that the distributions $P_N(s)$ on finite lattices
at $p=p_{c,N}$ fulfill roughly a scaling law
\be
   P_N(s) \sim 1/s
\ee
for $s<<N^2$, i.e. in the usual notation \cite{stauffer} we have
$\tau\approx 2$. Results for larger lattices and $p<p_c$ are shown
in fig.5. We see there again roughly $P(s) \sim 1/s$ for
$s<(p_c-p)^{-\gamma}$ (notice that bin sizes are
$\propto s$ in fig.5, in contrast to fixed bin sizes in fig.1). But
this is superimposed by oscillations whose amplitude increases as we
approach the critical point. The latter is very different from
the Ising ($q=2$) and percolation ($q=\infty$) limits, but was also
observed for $q=4$ and for the time distribution for $q=3$ and $q=4$.
On the other hand, model B leads to cluster size distributions which
are qualitatively as in percolation and in the Ising model, see fig.6.

Results for $p_c$ and for the critical exponent $\gamma$ are summarized
in table 1, for both model classes. For class A, we see that $p_c$
interpolates monotonically between the Ising and percolation limits. The
latter is approached very quickly. On
the other hand, $\gamma$ seems first to fall below the Ising value 7/4,
and rises only very slowly towards the percolation limit $43/18=2.388$
for $q\to\infty$.
For class B we see that there is a phase transition for all values of q
with $p_c(q) \to 1$ very slowly for $q\to\infty$. The amplitudes in the
scaling laws for class B seem to tend towards zero for $q\to\infty$
(the clusters become very small except when $p$ is very near $p_c$),
in agreement with the arguments given above. We do not quote exponents
$\delta$ because they are very strongly affected by corrections to
scaling, and we do not quote exponents $\tau$ for class B for the same
reason.

Superficially, our models resemble the Potts model. There, one has
a first order transition for $q>4$. One might thus wonder whether there
is some range in $q$ in which one of our models has also a first order
transition. We have not found any indication for that.

In summary, we have in this paper presented evidence for two new classes
of non-equilibrium critical phenomena. Both classes use Potts spins,
and are defined via cluster evolution rules similar to those of the
Wolff dynamics for the Potts model. The most interesting aspect of the
first class is that it interpolates between the Ising model ($q=2$)
and percolation ($q=\infty$). For this class we have not been able
to find any observable which shows broken detailed balance. We thus
cannot exclude that there exists for this model also an equilibrium
version, possibly with long range interactions (otherwise it would
be hard to understand why these models have not been found before).

In class B, breaking of detailed balance is evident. Indeed, the
transition in class B is from desynchronized cyclic behavior to
synchronized one by formation of a percolating cluster of sites with
common phase. Synchronization in spatially extended models was
studied recently in \cite{bennett} and \cite{grinstein}. In
\cite{bennett} it was argued that no synchronization can occur in
stochastic discrete models with $>2$ states. Continuous models were
studied in \cite{grinstein} where a connection was established with surface
roughening in Kardar-Parisi-Zhang growth. The latter suggests that no
phase transition with long range order occurs in $d=2$ for such models
either. These predictions are in clear contrast to our findings. This
reflects basic differences in the dynamics underlying model B and
the systems studied in \cite{bennett,grinstein}. While it is assumed
in the latter that the phase proceeds rather uniformly with locally
independent fluctuations, the phase progress in the ordered phase of
model B is dominated by rigid flippings of entire large clusters.

Finally, we have neither done any simulations in $d>2$, nor have we
attempted a field theoretic treatment. Using methods of
\cite{doi,scheun,peliti} it should not be difficult to set up the
field theory. But it might be much harder to find a systematic
perturbation expansion.

Part of this work was done when one of us (P.G.) was visiting Rome
University. He wants to thank A. Vulpiani and G. Paladin for their
generous hospitality and for numerous enlightening discussions.

\vspace{3.cm}
\eject

{\bf Table 1:}

\vspace{.2cm}

\begin{tabular}{|r|c|c|c|c|c|} \hline

           &  $q=2$  &  $q=3$  &  $q=4$  &  $q=8$  & $q=\infty$ \\ \hline
model A:   &         &         &         &         &            \\
     $p_c$ &  0.5858 &  0.5330 &  0.5170 &  0.5029 &  0.5       \\
  $\gamma$ &  1.75   &  1.65   &   1.79  &  1.88   &  2.388     \\
  $\tau$   &  2      &$\approx 2$&$\approx 2$&$\approx 2$& 2.055 \\ \hline
model B:   &         &         &         &         &            \\
     $p_c$ &  0.5858 &  0.633  &  0.6667 &  0.7393 &  1         \\
  $\gamma$ &  1.75   &  1.35   &   1.21  &  0.71   &  --        \\ \hline
\end{tabular}

\vspace{2.cm}

\eject

\vspace{1.3cm}

\section*{Figure Caption:}

{\bf Fig.1:} Cluster size distributions for $q=3$, lattice size $N =
    1024$, and 3 different values of $p$: $p=0.53175 (\Diamond), p=0.532$
    (boxes), and $p=0.53225 (\triangle)$. Data were binned into bins
    of size $\Delta s = 10^4$. The straight line has slope $-0.9$.

{\bf Fig.2:} Finite size critical values $p_{c,N}$ versus $1/N$ for
    $q=3$. For large $N$ the data seem to follow a straight line, the
    extrapolation of which to $1/N = 0$ gives $p_c$ (dashed line).

{\bf Fig.3:} Log-log plot of average cluster size for $q=3$ against $p-
    p_0$, for $p_0 = 0.5325,0.533$ and 0.5335. A roughly straight line is
    seen for $p_0=p_c\approx 0.533$, but notice that there are small
    but significant deviations.

{\bf Fig.4:} Same as fig.3, but for the average cluster life time.

{\bf Fig.5:} Cluster size distributions ($q=3$) for subcritical
    $p$ on practically infinite lattices. Values of $p$ are 0.5292,
    0.531, and 0.5318. Data are binned into bins $[s,2s-1]$.

{\bf Fig.6:} Similar as fig.5, but for model B with $q=3$. Values of $p$
    are 0.61, 0.63, and 0.6326.

\end{document}